\documentclass[prl,twocolumn,showpacs,floatfix,preprintnumbers,amsmath,amssymb,superscriptaddress]{revtex4}

\usepackage{graphicx}
\usepackage{dcolumn}
\usepackage{bm}
\usepackage{color}
\usepackage{color}
\usepackage[latin1]{inputenc}
\usepackage[urlcolor=blue]{hyperref}
\hypersetup{backref, colorlinks=true, linkcolor=blue, citecolor=blue}

\begin{document}

\title{{Intra-Atomic and Local Exchange Fields in the Van der Waals Magnet CrI$_{3}$}}

\author{Anirudha Ghosh}
\affiliation{Uppsala University, Department of Physics and Astronomy, Box 516, SE-751 20 Uppsala, Sweden}

\author{H. Johan M. J\"onsson}
\affiliation{Uppsala University, Department of Physics and Astronomy, Box 516, SE-751 20 Uppsala, Sweden}

\author{D. J. Mukkattukavil}
\affiliation{Uppsala University, Department of Physics and Astronomy, Box 516, SE-751 20 Uppsala, Sweden}

\author{Y. Kvashnin}
\affiliation{Uppsala University, Department of Physics and Astronomy, Box 516, SE-751 20 Uppsala, Sweden}

\author{D. Phuyal}
\affiliation{Department of Applied Physics, KTH Royal Institute of Technology, SE-106 91 Stockholm, Sweden}


\author{M. Agåker}
\affiliation{Uppsala University, Department of Physics and Astronomy, Box 516, SE-751 20 Uppsala, Sweden}
\affiliation{MAX IV Laboratory, Lund University, PO Box 118, SE-22100 Lund, Sweden}

\author{Alessandro Nicolaou}
\affiliation{Synchrotron SOLEIL, L'Orme des Merisiers, Saint-Aubin, BP48, F-91192 Gif-sur-Yvette, France}

\author{M. Jonak}
\affiliation{Kirchhoff Institute of Physics, Heidelberg University, Heidelberg, Germany}

\author{R. Klingeler}
\affiliation{Kirchhoff Institute of Physics, Heidelberg University, Heidelberg, Germany}

\author{M. V. Kamalakar}
\affiliation{Uppsala University, Department of Physics and Asronomy, Box 516, SE-751 20 Uppsala, Sweden}

\author{Håkan Rensmo}
\affiliation{Uppsala University, Department of Physics and Astronomy, Box 516, SE-751 20 Uppsala, Sweden}

\author{Tapati Sarkar}
\affiliation{Department of Materials Science and Engineering, Box 35, Uppsala University, SE-751 03, Sweden}

\author{Alexander N. Vasiliev}
\affiliation{National University of Science and Technology "MISiS", Moscow, 119049, Russia}
\affiliation{7Lomonosov Moscow State University, Moscow 119991, Russia}
\affiliation{Ural Federal University, Ekaterinburg 620002, Russia}

\author{Sergei Butorin}
\affiliation{Uppsala University, Department of Physics and Astronomy, Box 516, SE-751 20 Uppsala, Sweden}

\author{J.-E. Rubensson}
\affiliation{Uppsala University, Department of Physics and Astronomy, Box 516, SE-751 20 Uppsala, Sweden}

\author{Olle Eriksson}
\affiliation{Uppsala University, Department of Physics and Astronomy, Box 516, SE-751 20 Uppsala, Sweden}
\affiliation{School of Science and Technology, \"Orebro University, SE-701 82 \"Orebro, Sweden}

\author{Mahmoud Abdel-Hafiez}
\affiliation{Uppsala University, Department of Physics and Astronomy, Box 516, SE-751 20 Uppsala, Sweden}

\date{\today}

\begin{abstract}
We report on a combined experimental and theoretical study on CrI$_{3}$ single crystals by employing the polarization dependence of resonant inelastic X-ray scattering (RIXS). Our investigations reveal multiple Cr 3$d$ orbital splitting ($dd$ excitations) as well as magnetic dichroism (MD) in the RIXS spectra which is evidence of spin-flip in the $dd$ excitation. Interestingly, the $dd$ excitation energies are similar on both sides of the ferromagnetic transition temperature, $T_{C} \sim$  61\,K, although MD in RIXS is predominant at 0.4 tesla magnetic field below $T_{C}$. This demonstrates that the ferromagnetic superexchange interaction that is responsible for the intra-atomic exchange field, is vanishingly small compared to local exchange field that comes from exchange and correlation interaction among the interacting Cr 3$d$ orbitals. The investigation presented here demonstrate that the electronic structure of bulk CrI$_{3}$ is complex in the sense that dynamical electron correlations are significant. The recorded RIXS spectra reported here reveal clearly resolved Cr 3$d$ intra-orbital $dd$ excitations that represent transitions between electronic levels that are heavily influenced by multi-configuration effects. Our calculations employing the crystal field TTmultiplet theory taking into account the Cr 3$d$ hybridization with the ligand valence states and the full multiplet structure due to intra-atomic and crystal field interactions in O$_{h}$ and D$_{3d}$ symmetry, clearly reproduced the dichroic trend in experimental RIXS spectra.
\end{abstract}


\maketitle


Over the last few years, two-dimensional (2D) van der Waals materials have spurred enormous interest due to their unique magnetic properties and potential to develop multi-functional electronics and spintronics devices \cite{1,2,3,4,5}. Among various 2D systems, the family of CrX$_{3}$ (X = Cl, Br, and I) has recently been at the center of widespread research in developing high-quality samples, from monolayer to bulk single crystals. Although ferromagnetism in semiconducting bulk CrI$_{3}$ is known since the mid-twentieth century \cite{6}, recent experiments reveal layer-dependent magnetic phases \cite{7,8}: from ferromagnetism in the monolayer to antiferromagnetism in the bilayer and back to ferromagnetism in the trilayer and bulk. This has been postulated to arise from competing interlayer and intra-layer antiferromagnetic and ferromagnetic interactions, respectively \cite{8}. According to Mermin-Wagner theory \cite{9}, in absence of intrinsic anisotropy, long-range magnetic order is strongly suppressed in a 2D isotropic Heisenberg system due to spin fluctuation at a finite temperature. However, any deviation from a pure Heisenberg interaction, such as the influence of the magnetocrystalline anisotropy (MCA), would be able to stabilize a magnetically ordered state, which is most likely what happens in CrI$_{3}$ \cite{10}. In CrI$_{3}$, the MCA energy is $\Delta E_{MCA}$ $\sim$ 0.5\,meV \cite{11}, a large value whose origin has been a topic of enormous interest in the past couple of years with continued debate. X-ray magnetic circular dichroism (XMCD) studies at Cr $L_{2,3}$ and I $M_{4,5}$-edges revealed a strongly suppressed single-ion anisotropies of Cr 3$d$ and I 5$p$ states, whose energies are too small to explain the origin of $\Delta E_{MCA}$. Despite this, there is a considerable spin-orbit coupling (SOC) strength ($\sim$ 0.63\,eV) in the I 5p state and strong Cr 3$d$-I 5$p$ hybridization \cite{12,13}. The I-5$p$ state SOC strength primarily contributes to the MCA (in contrast to the much smaller SOC of Cr 3$d$ $\sim$ 0.05\,eV), either through the Cr 3$d$-I 5$p$-Cr 3$d$ superexchange hopping or via substantial admixture of band states \cite{14}. Superexchange in this material is stabilized by the inherent CrI$_{3}$ crystal structure constituting of trivalent Cr ions in an octahedral environment of I ions and the in-plane Cr-I-Cr bond angle $\sim$ 95$^{\circ}$. In this configuration, according to Goodenough-Kanamori-Anderson (GKA) rule, the Cr-Cr coupling is expected to be primarily ferromagnetic \cite{15}. Therefore, the origin of long-range ferromagnetism could be attributed to the high SOC of I 5$p$ states mediating in the anisotropic Cr 3$d$-I 5$p$-Cr 3$d$ superexchange hopping through 95$^{\circ}$ in-plane Cr-I-Cr bond angle \cite{16}. Recent theoretical results suggest that the magnetic interaction is not purely ferromagnetic, instead there is a competition between the ferromagnetic and antiferromagnetic interaction, which involve different Cr 3$d$ orbitals \cite{10,11,17}. Calculations also suggest a sizable contribution of the orbital resolved components of Cr 3$d$ states from nearest neighbor interactions \cite{18,19}. In monolayer CrI$_{3}$, recent electronic structure calculations reveal substantial exchange splitting of the spin-polarized Cr 3$d$ $t_{2g}$ and $e_{g}$ states due to ferromagnetic Cr-I-Cr superexchange interaction \cite{19}. However, the electronic structure of this (and similar) materials is complex, with expected competition from kinematic (band formation) effects and on-site Coulomb repulsion (e.g. as parametrized by the Hubbard U). It is expected that the Cr 3$d$ inter-orbital interactions mediated by the I 5$p$ orbital hold the key to understanding most electronic, magnetic, and transport properties of CrI$_{3}$. Therefore, probing the Cr 3$d$ orbital is indispensable for an in-depth understanding of the magnetic properties of CrI$_{3}$. To directly access the Cr 3$d$ orbitals, we carried out resonant inelastic x-ray scattering (RIXS) studies at the Cr $L_{2,3}$-edge of bulk CrI$_{3}$ single crystal sample with circularly polarized x-rays. Soft x-ray RIXS, corresponding to the direct transition from Cr 2$p$ core-levels to the Cr 3$d$ orbitals, allows us to study orbital ($dd$) excitations. In addition, magnetic circular dichroism (MCD) in RIXS enabled an in-depth understanding of MCD in the $dd$ excitation which is sensitive to the symmetry of the d orbitals. The experimental investigations are found to be well corroborated by electronic structure calculations using DMFT level of approximation.

\begin{figure}[tbp]
\includegraphics[width=20pc,clip]{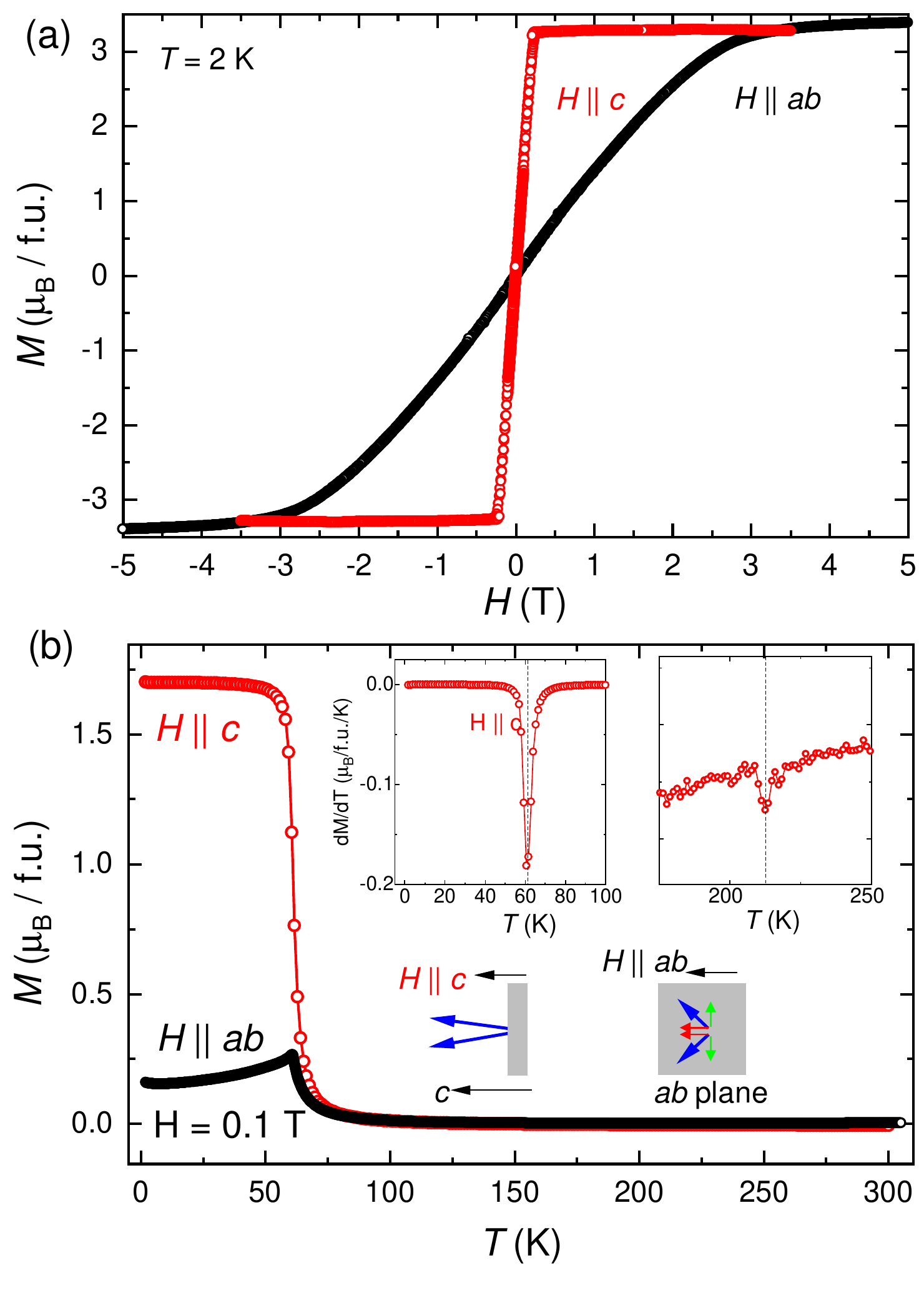}
\caption{(a) $M$ ($H$) measured at 2\,K, and (b) $M$($T$) plots with applied dc magnetic field parallel to the $c$-axis and $ab$-plane. Top Insets in (b) show the d$M$/d$T$ plots to extract $T_{C}$ and $T_{S}$. At an in-plane magnetic field ($H \parallel ab$: $H$=0.1\,T), the magnetization curve in (b) resembles in-plane anti-ferromagnetic interaction}
\end{figure}

Figure 1(a) shows the magnetization ($M$) vs magnetic field plots at 2\,K and at two alignment of the magnetic field, $H \parallel c$, and $H \parallel ab$. Although the saturation magnetization (MS) along both directions are similar, there is a substantial difference in the saturation fields for $H \parallel c$ ($H^{s}_{c}$) and $H \parallel ab$ ($H^{s}_{ab}$). Therefore, the anisotropy field is given by  $\Delta H_{s}$ = $H^{s}_{ab}$ - $H^{s}_{c}$ $\sim$ 3.16\,T, from which we deduce a magnetic anisotropy energy of $\sim$ 0.55\,meV, in agreement with a previous report\cite{11}. Figure 1(b) shows $M$ vs. temperature (MT) plots at an external magnetic field of 0.1 T applied parallel to the c-axis and the ab plane. As opposed to the saturating MT behavior below TC for $H \parallel c$, the MT tends to decrease below $T_{C}$ for $H \parallel ab$. This is reminiscent of in-plane antiferromagnetic (AFM) spin orientation. The out-of-plane (easy axis) spins will not completely polarize along the $ab$-plane at a nominal field of 0.1\,T due to MCA (see Fig 1(a); a magnetic field of more than 3\,T is required to overcome the MCA energy and polarize all the spins along the ab-plane). However, the projections of the canted c-axis spins on the $ab$-plane, at $H \parallel ab$ 0.1\,T could be AFM (bottom inset in Fig.\,1(b)) which explains the MT line-shape for $H \parallel ab$. Also, since these in-plane spins are not collinear at a nominal magnetic field (0.1\,T), which explains why at the lowest measured temperature (2\,K) there is still a small but finite $M$. The temperature derivative of M (inset of Fig.\,1b) gives a sharp peak at 61\,K and a weak peak near 212.5\,K, which are attributed to the FM to paramagnetic (PM) transition temperature ($T_{c}$) and phase transition from rhombohedral (R3) to monoclinic (C2/m) crystal structure ($T_{S}$), respectively. The results shown in Fig.\,1 are in agreement with our recent study \cite{20} and other reports \cite{21,22,23}.

\begin{figure}[tbp]
\includegraphics[width=21pc,clip]{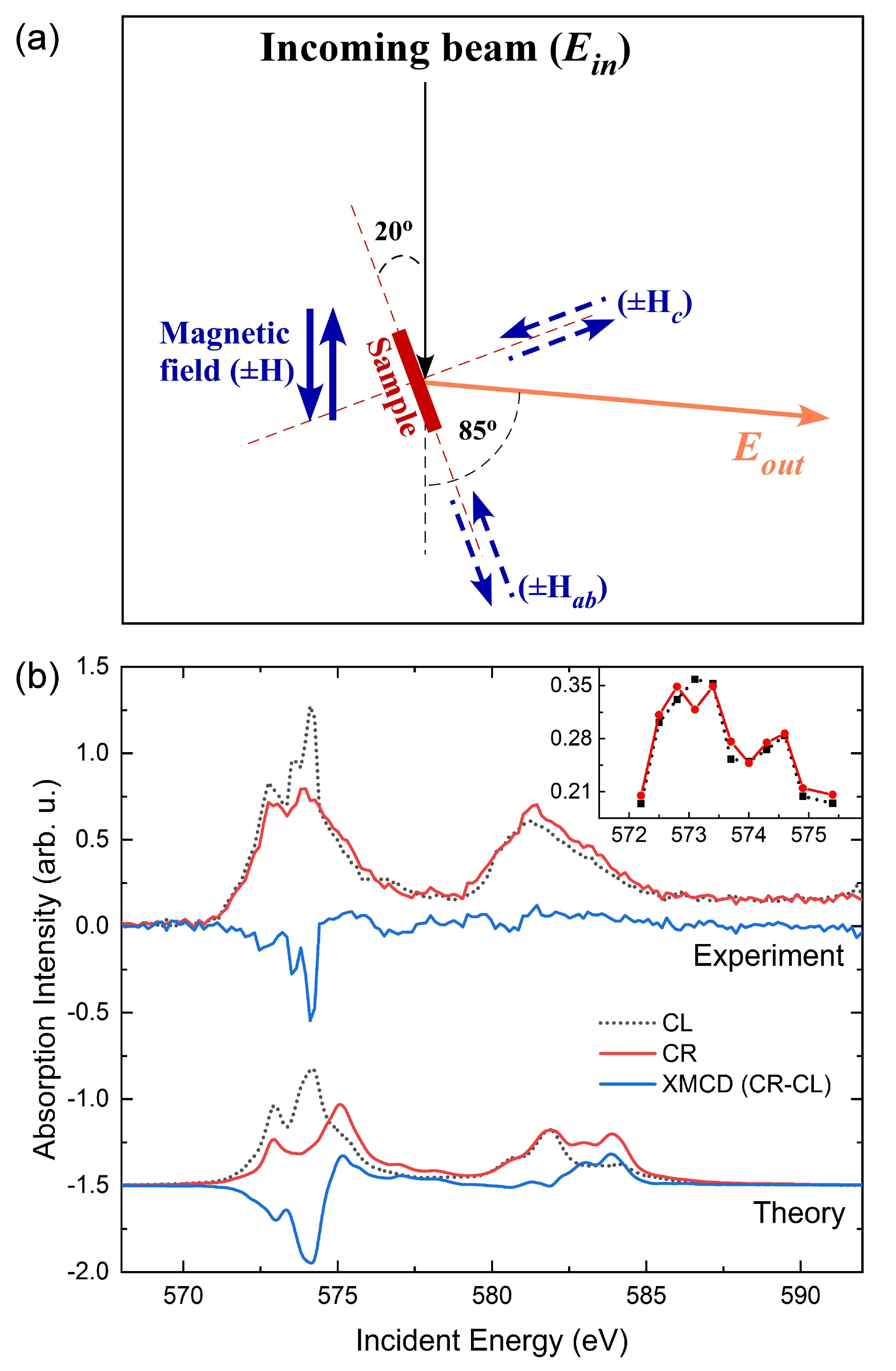}
\caption{(a) Schematic illustration of experimental geometry. The sample was illuminated with circularly polarized x-ray photons of energy $E_{in}$ at 20$^{\circ}$ grazing to the sample surface. The scattered rays ($E_{out}$) were detected by the spectrometer fixed at 85$^{\circ}$ scattering angle. (b) (top) Cr $L_{3,2}$ XAS and XMCD spectra measured at 20\,K and +$H$ magnetic field. (bottom) Calculated XAS and XMCD spectra. Inset shows the area under the energy loss region between 0.6 and 5\,eV of the RIXS spectra}
\end{figure}

To probe the in-plane FM interactions at a nominal in-plane magnetic field (above the $H \parallel c$ saturation field), we have carried out X-ray magnetic circular dichroism (XMCD) and RIXS studies at the Cr $L_{2,3}$ edge. Cr $L_{2,3}$-edge RIXS measurements were carried out at the SEXTANTS beamline\cite{24,25} of SOLEIL Synchrotron using the novel MAGELEC sample environment \cite{26}. Left and right circular (CL and CR) polarized X-rays of energy across the Cr $L_{2,3}$-edge ($E_{in}$) were focused at 20$^{\circ}$ grazing to the sample surface ($ab$ plane). The scattering angle was set to 85$^{\circ}$ and the overall energy resolution to 200\,meV. As presented in figure 2 (a) a static magnetic field of 0.45 T was applied parallel to the incoming beam and tilted by 20$^{\circ}$ with respect to the sample surface. This means that the components of the magnetic field parallel to the $ab$-plane ($H_{ab}$) and $c$-axis ($H_{c}$) were 0.36 and 0.14\,T, respectively. Note that although $H_{ab}$ is not enough to completely polarize the Cr spins in the $ab$ plane (3\,T magnetic field is required, Fig.\,1(a)), $H_{c}$ may be high enough to completely saturate the Cr spins along the c-axis. Therefore, the measured RIXS spectra (Fig.\,3) originate primarily from the saturated spins along the c-axis.  The XMCD presented in Fig.\,2(b) was obtained with H parallel to the c-axis. Measurements were carried out at 20 and 100\,K, i.e. below and above the Curie temperature. The crystal is in the rhombohedral phase at both these temperatures. Figure 2(b) shows the Cr $L_{2,3}$ edge x-ray absorption spectra (XAS) of CrI$_{3}$ single crystal sample at 20\,K (ferromagnetic phase of rhombohedral (R3) crystal structure). XAS was recorded in total electron yield mode with CL and CR polarized x-rays. XMCD is the difference between the CR and CL polarized XAS as shown in the top of Fig.\,2(a). Calculations involve using the multiplet of the ground state configuration together with corresponding quantum numbers (including S, L and J) using two different methods (Quanty\cite{27} and Impurity Model \cite{28}; see SI for details of calculations), both adopting the Anderson impurity model (with spins aligned parallel to the magnetic field, along the $c$-axis). The two approaches are very similar, with the main difference being the connection to parameters from bulk calculations on DFT or DMFT level. The results of the two methods are very similar, and for this reason we show in Fig.\,2(b) only results of Quanty. Here the multiplet of the ground state configuration was calculated by TT Multiplets taking into account the 3$d$ spin-orbit interaction in C3i symmetry. The calculated XAS and XMCD are shown in the bottom of Fig.\,2(b) which agree fairly well with the experimental data. We also note that from the DMFT calculations we obtain a Cr magnetic moment of 3.24 Bohr magnetons per atom, which is close to the experimental value and distinctly different for the results of LDA+U calculations (3.70 Bohr per atom).

\begin{figure*}[!t]
\includegraphics[width=1.8\columnwidth]{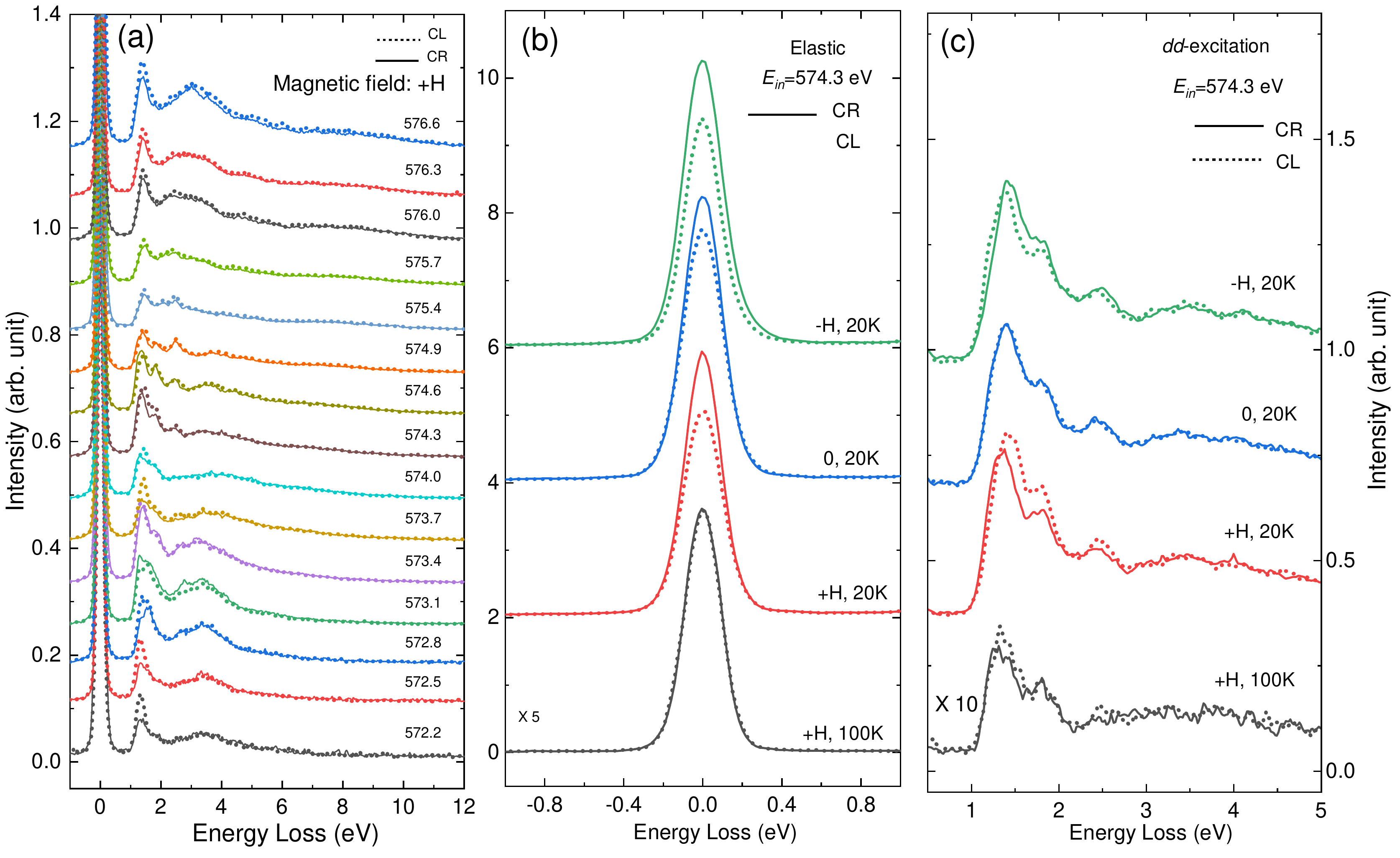}
\caption{(a) RIXS energy loss spectra at 20\,K at various circularly polarized (CL and CR) excitation energies. +H was the direction of the magnetic field. (b) MD in the elastic (left) and $dd$-excitation (right) features at 20\,K (-$H$, 0 and +$H$ magnetic fields) and 100\,K (+$H$ magnetic field).}
\label{fig:calc}
\end{figure*}

Figure\,3(a) shows RIXS energy loss spectra at various excitation energies (for CR and CL polarized x-rays) mentioned alongside each spectrum. In a conventional RIXS process, a monochromatic beam of energy $E_{in}$ is incident onto the sample, and one measures the energy distribution of the emitted radiation (Eout) converted into an energy loss ($E_{in}$ - $E_{out}$) scale to determine any low-energy excitations of the system. The energy loss at 0\,eV corresponds to the elastically scattered photons (where $E_{in}$ = $E_{out}$), while other features correspond to $dd$ excitations, charge transfer (CT), and Cr $L_{\alpha}$ fluorescence, in the order of increasing energy loss. The energy loss position of some of these features like $dd$ and CT excitations [28] are independent of incident energy and are also known as Raman-like losses. We identify the $dd$ excitation features (1.45$\pm$0.02, 1.75$\pm$0.02, 2.50$\pm$0.05\,eV) between 1-3\,eV energy loss as shown in Fig.\,3(a) and for better clarity we re-plot the RIXS spectrum at the resonance L3 energy (574.3\,eV) in Fig.\,3(b). Above 575\,eV excitation energy (Fig.\,3(a)), the broad feature (between 2-5\,eV) resembles the emission ($L_{\alpha}$) from Cr $3d_{5/2}$ to $2p_{3/2}$ in addition to the buried $dd$ and CT features. The broad feature in the RIXS spectra excited near the absorption threshold (near 572\,eV) signifies the metal-ligand CT (3.50$\pm$0.10\,eV), which gets smeared out in intensity relative to the $dd$ excitation features, near the resonant energy. The difference in intensities between the CL and CR RIXS features at 20\,K, in particular at $dd$ excitations, is clearly visible at resonant excitation and also at other excitation energies in Fig.\,3(a). The integrated intensities between the regions 0.6 and 5\,eV, that correspond to the energy loss region, of CL and CR features for several excitation energies are plotted in the inset of Fig.\,2(b). A clear contrast in the integrated intensities can also be observed in addition to the XMCD contrast in Fig.\,2. Upon reversing the direction of the magnetic field (to -$H$), the CL and CR RIXS features in the $dd$ excitation also switch their relative intensity depending on the light helicity, and at zero magnetic field, their intensities are similar (Fig.\,3(b)). This behavior is direct evidence of magnetic dichroism (MD) in RIXS and is not an experimental artifact.
\begin{figure*}[!t]
\includegraphics[width=1.8\columnwidth]{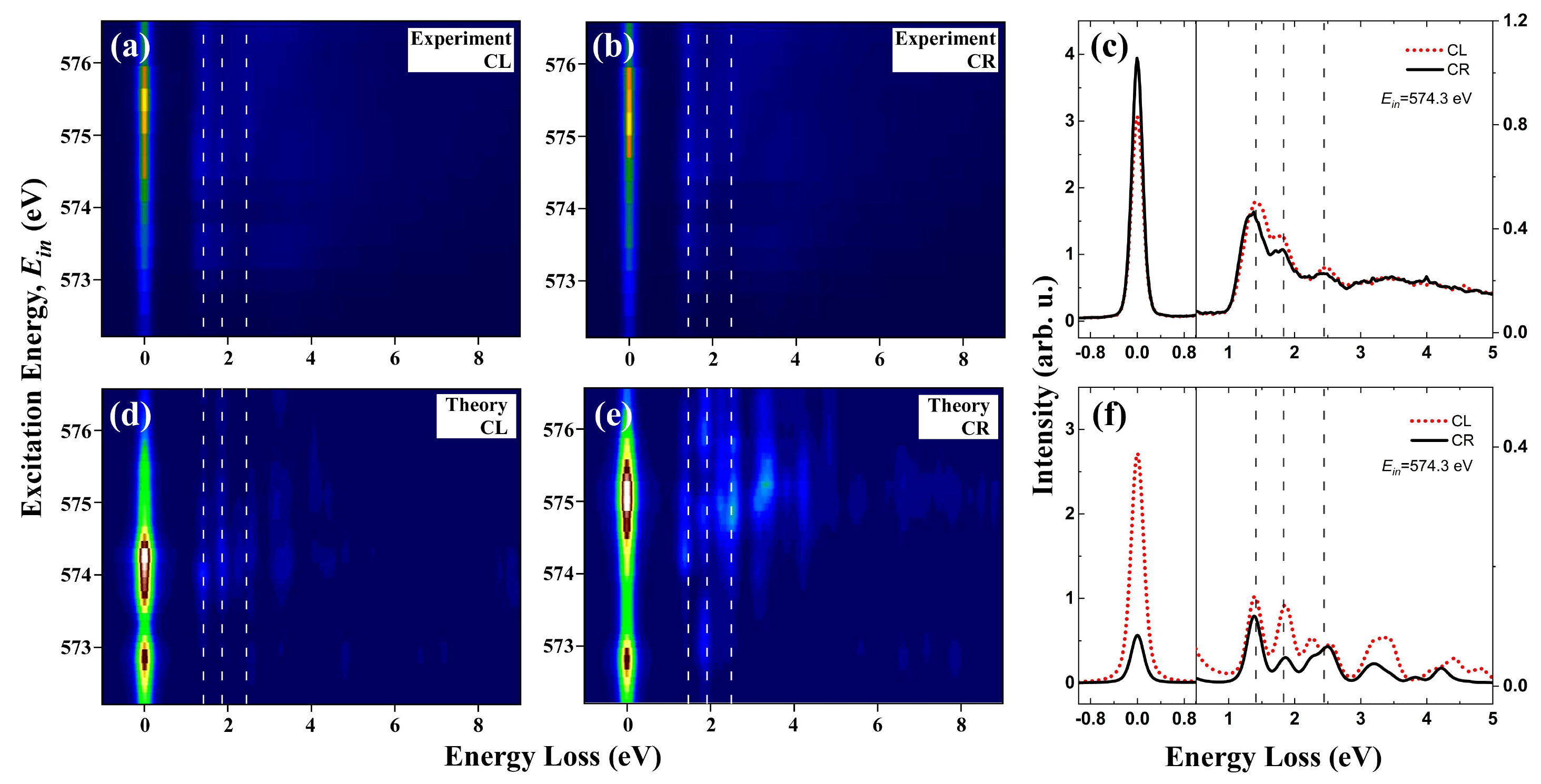}
\caption{Experimental RIXS maps measured at 20\,K with (a) CL and (b) CR polarized x-rays incident grazing to the sample surface and a magnetic field along +$H$ direction. (c) dichroic RIXS spectra at resonant excitation along +$H$ direction. Calculated RIXS maps with (d) CL and (e) CR polarized x-rays. (f) line spectra at 574.3\,eV excitation energy obtained from (d) and (e).}
\label{fig:calc}
\end{figure*}
The MD in dd-excitation is evident of spin-flip excitation corresponding to $t_{2g}\rightarrow t_{2g}$ or $t_{2g}\rightarrow e_{g}$ in the final state. As discussed earlier in the present measurement geometry, the MD in RIXS is primarily due to the Cr spins parallel to the $c$-axis. This is due to the fact that even at $H_{c}$ = 0.14 T, the Cr spins parallel to the $c$-axis are fully ferromagnetically saturated, in contrast to unsaturated in-plane Cr spins at $H_{ab}$ = 0.36\,T (see Fig. 1(a)). However, at the nominal magnetic field, the in-plane spins can be anti-ferromagnetically oriented as already observed in Fig.\,1(b).A similar situation is present in $\alpha$-Fe$_{2}$O$_{3}$\cite{30} where the spin-flip excitations involving the $e_{g}$ states, were interpreted as the origin of anisotropic SOC, which was responsible for the Dzyaloshinskii-Moriya (DM) interaction. Although structurally Fe$_{2}$O$_{3}$ and CrI$_{3}$ are entirely different systems, DM interaction may also be present in CrI$_{3}$ which is believed to be the origin of the spin-gap at the Dirac point of spin-wave (SW) excitation spectra, observed in inelastic neutron scattering experiments\cite{31}. Several calculations\cite{19,31,32} showed that in the absence of DM interaction the origin of the spin-gap cannot be explained and the system would eventually be similar to a spin analogue of graphene (spin-gapless at the Dirac point). There is, however, a continued debate on the strength of DM interaction. A realistic estimation of DM interaction strength, by Kvashnin et al. \cite{19}, as compared to the Heisenberg exchange interaction, underestimates the spin-gap value, although it fairly accurately predicts the splitting in the Cr 3$d$ orbital and corresponding spin-flip $dd$ excitations (1.45, 1.75, and 2.50\,eV). Kitaev interaction\cite{33} and electron correlation\cite{34} may also contribute to the origin of the spin-gap in SW excitation. Recently, we found a possible evidence of a complex magnetic spin liquid-like state at high pressures and very low temperatures in CrI$_{3}$ single crystal\cite{20}. However, further investigations are required to quantify the roles of DM and Kitaev interactions, associated with the SOC strength, in accurately estimating the magnitude of the spin-gap and spin-flip $dd$ excitations.

In the paramagnetic phase of CrI$_{3}$ (with the same rhombohedral crystal structure as that at 20 K) [21], at 100 K and in the presence of the in-plane magnetic field ($H$ =0.4\,T), there can be residual net magnetic moment and therefore, we can see some MD in the $dd$ excitation, although it is absent in the elastic peak (Fig.\,3(b)). Importantly, we do not observe any significant peak shift or splitting between the RIXS spectra at 20\,K and 100\,K. This demonstrates that the ferromagnetic superexchange interaction that is responsible for the inter-atomic exchange field, is vanishingly small compared to the local exchange field that comes from intra-atomic exchange and correlation interaction among the interacting Cr 3$d$ orbitals of a given site. The latter interaction is driven by local exchange interactions of order eV, which is clearly much larger than estimates of inter-atomic exchange which is on the meV level. Figures 4(a) and (b) plot the experimental RIXS map near the Cr $L_{3}$ region for CL and CR polarization, respectively at 20\,K. RIXS spectra at the Cr $L_{3}$ resonance excitation are shown in in Fig.\,4(c). For comparison, we plot the corresponding calculated RIXS map in Fig.\,4(d) and (e) and the line spectra at resonance $L_{3}$ excitation in Fig.\,4(f). For best comparison with the theoretical line spectra, we replot, in Fig.\,4(c), the same spectra as shown in Fig.\,3(b) (+$H$), where the CCD was exposed for a total of 30 minutes. The dichroism and the energy loss peak positions in the calculated spectra, considering a SOC in the system, fairly agree with that of the experiment. However, without SOC, similar spectra were not reproduced to be consistent with the experimental one. Therefore, SOC plays a significant role and may also be responsible for the DM interaction when considering the projected in-plane FM spins in the in-plane AFM matrix at a nominal in-plane magnetic field. However, further calculations are necessary to explicitly understand and quantify the effects of DM and Kitaev interactions, which are based on SOC, on the orbital excitations in CrI$_{3}$ and related family of 2D materials.


To summarize, we have studied the Cr 3$d$ orbital excitations across the Curie temperature in bulk CrI$_{3}$ single-crystal sample. The RIXS spectra at both temperatures reveal Cr 3$d$ intra-orbital $dd$ excitations. The Cr-Cr inter-orbital interactions, via the I 5$p$ orbital, are responsible for the co-existence of FM and AFM interactions, of which $e_{g}$-$t_{2g}$ channel is the dominating one responsible for FM interaction\cite{19}. Despite the significant superexchange interactions and MCA, we could not observe exchange splitting of electronic states of Cr 3$d$ below $T_{C}$, which would result from inter-atomic exchange. This demonstrates that the intra-atomic exchange field is significantly stronger than that coming from the inter-atomic exchange. MD in RIXS spectra suggest spin-flip in the $dd$ excitation corresponding to $t_{2g}\rightarrow t_{2g}$ or $t_{2g}\rightarrow e_{g}$ in the final state, a possible ingredient to support DM interaction in CrI$_{3}$. Although both the DM and Kitaev interactions originate due to anisotropic SOC, further investigations are required to explicitly quantify their roles in the emergence of various exotic properties in CrI$_{3}$ and other honeycomb 2D systems. The experiments were complemented with theoretical calculations of XAS and XMCD, as well as ground state magnetic properties like the spin and orbital moments. Overall, we find that theory is in good agreement with the measured spectra from a DMFT level of approximation, when dynamical correlations are included. A static treatment of Coulomb interactions result in significant deviations from experiment, particularly as concerns the Cr spin-moment, (see supplemental information for more details). For the spectroscopic features we note that a theory on multi-configurational level reproduces experimental observations with high precision. The investigation presented here demonstrates that the electronic structure of bulk CrI$_{3}$ (and most likely of its 2D sister compounds) is complex in the sense that dynamical electron correlations and multi-configuration effects are significant. This is signaled both from the ground state (magnetic moment) as well as spectroscopic, excited state properties. The recorded RIXS spectra here reveal clearly resolved Cr 3$d$ intra-orbital $dd$ excitations that represent transitions between electronic levels heavily influenced by multi-configuration effects.

A.G. and MAH acknowledges financial support from Carl Tryggers Foundation and the Swedish Research Council (VR) under project No. 2018-05393. D. J. M. acknowledges support from SSF grant, RIF14-0064. O.E. acknowledge financial support from the Knut and Alice Wallenberg Foundation, eSSENCE, the Swedish Research Council (VR), the Foundation for Strategic Research (SSF). O.E, and P.T. acknowledge support from the ERC synergy grant (854843-FASTCORR). Y.K. acknowledges the support from VR (project No. 2019-03569) and Göran Gustafsson Foundation.  Support by the P220 program of Government of Russia through the project 075-15-2021-604 is acknowledged. The computations were enabled by resources provided by the Swedish National Infrastructure for Computing (SNIC) at the National Supercomputer Centre (NSC) partially funded by the Swedish Research Council through grant agreement no. 2018-05973. D.P. acknowledges the support from Swedish Research Council under Project No. 2020-00681. T. S. acknowledges financial support from the Swedish Research Council (VR Grant Nos. 2017-05030 and 2021-03675).

\end{document}